\documentclass[12pt]{article}
\usepackage{amsmath,amssymb}
\usepackage[dvips]{graphicx}
\usepackage{pdfpages}
\usepackage{pdflscape}
\usepackage{lmodern}
\begin{document}

\begin{center}

\section*{ The   quantum vortices dynamics: \\  spatio-temporal scale hierarchy\\  and origin of turbulence}

{S.V. TALALOV}

{Department of Applied Mathematics, Togliatti State University, \\ 14 Belorusskaya str.,
 Tolyatti, Samara region, 445020 Russia.\\
svt\_19@mail.ru}

\end{center}

\begin{abstract}
This study investigates the evolution and interaction of  quantum vortex loops with a small but non-zero radius  of  core ${\sf a}$. The quantization scheme of the  classical vortex  system is based on the approach proposed by the author  \cite{Tal,Tal_PhRF}.   
  We  consider  small  perturbations in  the  ring-shaped loops, which include   both helical-type shape variations and small excitations of the flow in the vortex core. 
The quantization of  the circulation $\Gamma$ is deduced from the  first principles of quantum theory. As a result of our approach, the set of quantized circulation values is wider than the standard one.
The developed theory introduces a hierarchical spatio-temporal scale in the quantum evolution of vortices. We also explore the applicability of this model for describing the origins of turbulence in quantum fluid flows. To achieve this specific objective, we employ the method of random Hamiltonians  to describe the interaction of  quantum vortex loops.
\end{abstract}


\vspace{7mm}

{\bf keywords:}   quantum vortices,  circulation quantization, quantum turbulence.
 
\vspace{5mm}

 {\bf PACS numbers:}   47.10.Df    ~~47.32.C

\vspace{5mm}

\section{Introduction}

~~~ The hypothesis that vortex interactions are fundamental in forming turbulent flow has long been suggested (see \cite{Feyn,Donn}, for example).
Moreover,  it has long been understood that
 the emergence of different spatial scales is an integral part of the turbulence \cite{Kolm}.  A significant number of works has been devoted to the study of this issue at the classical level, an overview of which can be found in the book \cite{Frisch}.
Despite this, the problem of constructing a theory describing the spatio-temporal hierarchy of a turbulent flow still exists \cite{PoMuKr_1}.
Addressing turbulence at the quantum level introduces new challenges \cite{Vinen} , as outlined in recent reviews and literature on the subject \cite{Tsu_1} and book   \cite{Sonin}.   
{A separate direction of research here is the numerical simulation of quantum turbulence. This area is actively developing  at the present \cite{GBBK}}

When studying complex systems at the quantum level,  the primary question is which variables to quantize and how to approach quantization. 
 Of course, in  simple systems\footnote{The author is referring to classical dynamical systems, which are usually considered in quantum mechanics textbooks.
At the same time, the  quantization of  such a simple canonical system as the harmonic oscillator in terms of action-angle variables demonstrates many unexpected and interesting results \cite{Kast}.}, such a question usually does not arise. In complex dynamical systems, the choice of variables for quantization is no longer that obvious. 
It can be stated that  Dirac’s perspective remains pertinent: 					{\it''\dots methods of quantization are all of the nature of practical rules, whose application depends on consideration of simplicity''} \cite{Dirac}.
{ As it seems, the quantum description of a single vortex is not a trivial task. 
In this study, we do not intend to make a detailed overview of this problem.  
We refer readers to references \cite{Donn} and \cite{Sonin} that  
 contain information on this issue for a large   time period.
 Among the recent studies, it is necessary to mention the works \cite{Sb1,Sb2}, where
description of quantum vortices in terms of the modified Navier-Stokes equation was considered.
It is also worth mentioning paper \cite{GladW}, in which a numerical analysis of the interaction of vortices is made within the framework of the Gross-Pitaevskii theory. }

This study is devoted to the description of  quantum evolution and a certain type interaction  of closed vortex loops.
We consider the simplest ring-type configurations: tangled and knotted loops \cite{Baren} are beyond the scope of this study.
From the authors' perspective,  the proposed approach   makes it possible to apply the quantum theory of many bodies  to  describe the interaction of vortex loops naturally.
The model  considered in this study builds upon the author's previous work \cite{Tal}.

{ The proposed approach is founded on the following fundamental assumptions. 
\begin{itemize}
\item  We  explore the  Hamiltonian dynamics of a small perturbed vortex ring  in Local Induction Approximation  in terms of new non-conventional Hamiltonian variables. We will discuss these points in detail in section 3 of the paper; 
\item We apply a group-theoretic approach to define  the full energy of a thin  (${\sf a} \approx 0$) vortex filament of  arbitrary radius R. Indeed,  the standard formula for the energy \cite{Saffm} leads to the unsatisfactory result for thin filaments:   the integral in the standard formula   diverges  for the filament with the core radius  ${\sf a} \to 0$. In order to implement this approach, we extend the space-time symmetry group of the theory to the central-extended Galilei group $\widetilde{\mathcal G}_3$. Lee algebra of the  group $\widetilde{\mathcal G}_3$ has three Cazimir functions:  
		 \[ {\hat C}_1 = \mu_0 {\hat I}\,,\quad 
  {\hat C}_2 = \left({\hat M}_i  - \sum_{k,j=x,y,z}\epsilon_{ijk}{\hat P}_j {\hat B}_k\right)^{\!2} 
  \quad {\hat C}_3 = \hat H -  \frac{1}{2\mu_0}\sum_{i=x,y,z}{\hat P}_i^{\,2}\,,\]                        
       where        ${\hat I}$ is the unit operator,     ${\hat M}_i$,   $\hat H$,  ${\hat P}_i$         and  ${\hat B}_i$  ($i = x,y,z$)
        are the respective generators of rotations, time and space translations and Galilean boosts, value $\mu_0$ is a central charge. 
		Traditionally, the function  ${\hat C}_3 $  can be interpreted as  an  ''internal energy of the particle'' as well as the central charge $\mu_0$ is usually interpretted as  ''mass''.  Therefore, we can define the energy as follows
		\begin{equation}
		\label{E_general}
 E = \frac{1}{2\mu_0}\sum_{i=x,y,z}{\bf p}^{\,2} + {\hat C}_3(j)\,,
\end{equation}
where ${\bf p}$ is vortex monentum an $j$ some set of the ''internal'' variables\footnote{In this paper, the issues related to the energy of the vortex ring are not considered.};
\item We quantize the considered dynamical systems and  calculate the spectra of the vortex energy and circulation.  Unlike the standard approach, quantum circulation values are derived from the first principles of quantum theory, but not postulated separately;
\item We deduce that  resulting spectra exhibit a fractal structure, which may better reflect turbulent flow behavior.
\end{itemize} }

One of the outcome of the proposed  method is a formula for quantized circulation values $\Gamma_n$.
In this study, we  expand the frame of our model. We  refine the formula for quantum circulation values and,  study various limiting cases. 
 The resulting formula differs from the standard one $\Gamma_n \propto  n$ (where  $n$ is an integer),  but leads to it in a certain  limiting  case.
In  general, we demonstrate the dependence of the numbers $\Gamma_n$ on the shape and size of the domain where the vortex ring evolves. Note that similar dependence is characteristic of quantum systems.
Moreover,  the set $\{\Gamma_n\}$ demonstrates the fractal structure in certain special cases of the proposed theory, potentially offering a better framework for turbulent flow than a series of natural numbers. 
As for the interaction of vortices, these issues were considered by the author in the papers
\cite{Tal_PhRF,Tal_PoF23}.  
We will not dwell on this in detail here.

\vspace{3mm}

\section{Dynamics of a perturbed vortex ring}

~~The starting point of our consideration is a single closed vortex filament that evolves in accordance with the local induction equation (LIE):

\begin{equation}
        \label{LIE_eq}
       \frac{ \partial {\boldsymbol{r}}({t^\ast} , S)}{\partial {t^\ast}} ~=~ \beta_1
       \frac{ \partial{\boldsymbol{r}}({t^\ast} ,S)}{\partial S}\times \frac{\partial^{\,2}{\boldsymbol{r}}({t^\ast},S)}{\partial S^2}\,.
        \end{equation}
 Vector function ${\boldsymbol{r}}(\cdot, S)$ means the the filament coordinates where symbol $S$ means  the natural parameter. 
This form of the equation LIE can be deduced  after 
replacing  natural time $t$ with the ''evolution parameter''   $t^\ast$:
\begin{equation}
        \label{ten_time}
			{t^\ast} ~=~ \frac{t\Gamma  }{4\pi}\,.	
	\end{equation}			
 It is known  that equation LIE  takes the form (\ref{LIE_eq}) due to a certain regularization with parameter $L$. As a consequence, the parameter $\beta_1 = \ln({2L}/{\sf a}) - 1$  appears in Eq.\ref{LIE_eq}. The value ${\sf a}$ is small but non-zero radius of the filament core. { The definition both auxiliary regularization parameter $L$ and other details regarding  the Eq. \ref{LIE_eq} can be found, for example, in the book \cite{AlKuOk}.}
 We assume  $\beta_1 = const$ for our  subsequent studies.

At the first stage, we  investigate small oscillations of a vortex ring of an arbitrary radius $R$. 
We assume that the inequality
\begin{equation}
        \label{R_f}
 R \ge R_f
 \end{equation}
holds for certain value $R_f$. This constant depends on the type of fluid we are considering.
For example,  the constant  $R_f$ may be related to the size of a fluid molecule, inter-atomic distance
or  size of a stable molecular cluster.
The oscillations that we intend to investigate  are caused by two processes.
First, we study  a small change in the initial circular shape of the vortex ring.
Second, we consider  the appearance of a weak flow in the core of the vortex filament.
This  flow leads to the additional terms   in Eq.\ref{LIE_eq}.
In our case, such additional terms are considered as small perturbations of the original equation.

Later in the paper, it will be more convenient  to use dimensionless parameters $\tau$ and $\xi$ instead of the evolution parameter   $t^\ast$ and the natural parameter $s$:
\begin{equation}
        \label{tau_s}
 \tau ~=~ \frac{t\Gamma  }{4\pi R^2}\,,	\qquad \quad \xi ~=~ \frac{S}{R}\,, \qquad
\xi \in [0,2\pi]\,.
 \end{equation}
Thus, we  consider the following equation for a projective vector 
${\mathfrak r} =  {\boldsymbol{r}}/R$:

\begin{eqnarray}
        \label{LIE_pert}
        \partial_\tau {\mathfrak r}(\tau ,\xi)  & = &
        \beta_1 \Bigl(\partial_\xi{\mathfrak r}(\tau ,\xi)\times\partial_\xi^{\,2}{\mathfrak r}(\tau ,\xi)\Bigr)   + \nonumber \\
				~~ & + & \varepsilon\omega\Bigl(2\,\partial_\xi^{\,3}{\mathfrak r}(\tau ,\xi) + 
        {3}\,\bigl\vert\, \partial_\xi^{\,2}{\mathfrak r}(\tau ,\xi)\bigr\vert^{\,2}\partial_\xi{\mathfrak r}(\tau ,\xi)\Bigr)\,,
				        \end{eqnarray}
where the value $\varepsilon$ is some small parameter and the value $\omega$ is a finite constant.
{ In total,  parameter $\varepsilon\omega$  is determined by the velocity of fluid flow within the vortex core.}
A detailed deducing  of this equation { as well as deducing the explicit expression of the parameter $\varepsilon\omega$  from the conventional physical values} was made in  the book \cite{AlKuOk}.
{ We consider the second term in Eq.(\ref{LIE_pert}) as a small perturbation, so in our case the explicit dependence is not important.}

For  convenience,  we use the representation of any smooth closed curve of length $S_0$   in the form
\begin{equation}
        \label{proj_r}
                                   {\mathfrak r}(\tau,\xi) ~=~  \frac{\boldsymbol{q}}{R} + 
          \int\limits_{0}^{2\pi}  \left[\, {\xi - \eta}\,\right] {\boldsymbol j}(\tau,\eta) d\eta\,, \qquad  R = \frac{S_0}{2\pi} \,,
                  \end{equation}
where vector ${\boldsymbol{q}} = {\boldsymbol{q}}(\tau)$ defines the position of the curve in the coordinate system (the   center of the circular vortex ring, for example).
The notation $[\,x\,]$ means the integer part of the number $x/{2\pi}$:
\begin{equation}
        \label{int_part}
 [\,0\,] = 0\,, \qquad  [\,x+2\pi\,] = [\,x\,]+1\,, \qquad \forall\,x\,.
\end{equation}
 Vector ${\boldsymbol j}(\tau,\eta)$ is the unit  tangent (affine) vector for this curve. Indeed, 
${\partial {\boldsymbol{r}}}/{\partial s} ~=~  {\boldsymbol j}$
in accordance with  formulas   (\ref{tau_s}),    (\ref{proj_r}) and   (\ref{int_part}).
The  components ${j}_k$ of the vector-function ${\boldsymbol j}(\tau,\xi)$  satisfy  to the conditions

            \begin{equation}
  \label{constr_j_0}
    \int\limits_{0}^{2\pi}{j}_k(\xi)\,d\xi  ~=~ 0\, \qquad  (k=x,y,z)\,.
   \end{equation}      
These conditions  ensure the closureness  of the curve in question: indeed,  the equality $ {\mathfrak r}(\tau,\xi + 2\pi) = {\mathfrak r}(\tau,\xi) $ will be fulfilled.
Since we are interested in solutions of the  equation (\ref{LIE_pert}) in the form of closed curves, we will use representation (\ref{proj_r}) for such solutions.

  For a circular vortex ring, the following solution of the Eq.(\ref{LIE_pert}) is valid for the case $\omega = 0\,$:	
\begin{eqnarray}
        \label{tang_v}
              {\boldsymbol{j}}(\xi) & = &  {\boldsymbol{j}_0}(\xi) ~\equiv~ \Bigl( - \sin(\phi_0 +\xi)\,, \quad \cos(\phi_0 +\xi)\,, \quad 0\Bigr)\,,\\[3mm]
						\label{q_def}	
			{\boldsymbol{q}}	& \equiv &  {\boldsymbol{q}_0}	 ~=~ 	{\boldsymbol{q}_0}(0) + \beta_1 R \tau \boldsymbol{e}_z\,,	
\end{eqnarray}
where the unit vector $\boldsymbol{e}_z$ defines the axis of the ring. In general, the quantities $\phi_0$ and ${\boldsymbol{q}_0}$ are the dynamical variables; we will return to the discussion of these variables later.  We also note that the parameter $\beta_1$ defines  the  finite velocity $\boldsymbol{u}_{v}$  of the ring in the space. 
The core radius   ${\sf a}$ is connected with the vortex ring velocity $\boldsymbol{u}_{v}$ (see, for example,  \cite{Saffm}):
\[ |\boldsymbol{u}_{v}|  ~\simeq~  \frac{\Gamma}{4\pi R}\ln\frac{8R}{{\sf a}}\,, \qquad {\sf a} << R\,.\]
Therefore, we  consider the rings with small but non-zero value of the core radius ${\sf a}$.


Let us consider the small perturbation   of the tangent vector (\ref{tang_v}): 
 
		\begin{equation}
        \label{tang_v_per}          																
		\boldsymbol{j}_0(\tau,\xi)  ~\to~ {\boldsymbol{j}_0}(\xi) + \varepsilon {\boldsymbol{j}}_p(\tau,\xi) \,,	\qquad 		\varepsilon << 1\,.													
\end{equation}
We assume that  introduced vector-function ${\boldsymbol{j}}_p(\tau,\xi)$ is a single-valued function of the parameter $\xi\in [0,2\pi]$.
Additionally, we  suppose that the 
identity 

\begin{equation}
        \label{iden_1}
				{\boldsymbol j}_p(\tau ,\xi){\boldsymbol{j}_0}(\xi) ~\equiv~ 0\,
	\end{equation}
is fulfilled for the perturbation amplitude  ${\boldsymbol{j}}_p(\tau,\xi)$.
{ Notation  ${\boldsymbol j}_p{\boldsymbol{j}_0}$ means the inner product of the vectors.}
Therefore, we  study the transverse excitations of the vector-function ${\boldsymbol{j}_0}(\xi)$ only. Some discussions in favor of this possibility can be found in the work \cite{Tal}. 
Therefore,

\begin{equation}
        \label{j-two}
{\boldsymbol j}_p(\tau ,\xi)   
  ~=~   j_\rho(\tau,\xi){\boldsymbol{e}}_\rho  +       j_z(\tau ,\xi)  {\boldsymbol{e}_z}\,,
\end{equation}
where the vectors $\{ {\boldsymbol{e}_\rho}\,,  {\boldsymbol{e}_\xi} \equiv {\boldsymbol{j}_0}\,, {\boldsymbol{e}_z}\,\}$ denote the local basis of the cylindrical coordinate system
$(\rho, \xi, z)$.   This system is connected with the  vortex ring in question.

The small perturbation of the circular vortex filament  without the restriction (\ref{iden_1}) has been studied in the work \cite{Ricc}. Among other results,   the appearance of knotted configurations of vortex filaments was demonstrated in the paper \cite{Ricc}.   As it was noted above, such objects are outside of  our consideration. Due to the simplification made, the subsequent quantization of the model will be simpler. 
We also should mention that small perturbations of straight vortex filaments were studied in the work \cite{Majda}.
Various aspects of the theory related to vortex ring oscillations have also been studied in the works \cite{Kop_Chern,Kik_Mam}

Perturbed  components of projective vector 
${\mathfrak r}$ satisfy the Eq.(\ref{LIE_pert}) for the case  $\varepsilon \not= 0$,
with the same $\varepsilon$ value   as in Eq.(\ref{tang_v_per}). 
It is well known that  the tangent vector ${\boldsymbol j}$  satisfies equation  for the continuous Heisenberg spin chain if the vector ${\boldsymbol r}$  satisfies equation (\ref{LIE_eq}):

\begin{equation}
        \label{CHSCeq}
        \partial_\tau {\boldsymbol{j}}(\tau ,\xi) ~=~ 
        \beta_1\,{\boldsymbol{j}}(\tau ,\xi)\times\partial_\xi^{\,2}{\boldsymbol{j}}(\tau,\xi)\,.
        \end{equation}
				The corresponding equation for the perturbation amplitude  ${\boldsymbol{j}}_p(\tau,\xi)$ is the linear equation. This equation is as follows:

\begin{equation}
        \label{lin_eq}
        \partial_\tau {\boldsymbol{j}_p}(\tau ,\xi) ~=~ \beta_1\, {\boldsymbol{j}_0}(\xi) \times 
				\left[\, {\boldsymbol{j}_p}(\tau ,\xi)  +   \partial_\xi^{\,2}{\boldsymbol{j}_p}(\tau,\xi)\, \right]\,.
				        \end{equation}

This case has been investigated in the paper \cite{Tal} in detail.
Now we consider the Eq.(\ref{LIE_pert}) for the case
 $\varepsilon\omega \not= 0$. As a consequence, the 
equation for the perturbation amplitude  ${\boldsymbol{j}}_p(\tau,\xi)$ gets additional term:

\begin{eqnarray}
        \label{lin_eq_1}
        \partial_\tau {\boldsymbol{j}_p}(\tau ,\xi) & = & \beta_1\, {\boldsymbol{j}_0}(\xi) \times 
				\left[\, {\boldsymbol{j}_p}(\tau ,\xi)  +   \partial_\xi^{\,2}{\boldsymbol{j}_p}(\tau,\xi)\, \right] + \\[2mm]
				~~ & + & \omega\, \bigl[2\,\partial_\xi^{\,3}{\boldsymbol{j}_0}(\xi) + 3\,\partial_\xi
					 {\boldsymbol{j}_0}(\xi)  \bigr]\,.\nonumber
				        \end{eqnarray}
In a standard way, we did not take into account terms of order $\varepsilon^2$ and higher.

Let us  introduce the notation ${\mathfrak J}$  for the complex perturbation amplitude:
\[  {\mathfrak J}(\tau ,\xi)  ~=~  j_\rho (\tau ,\xi) + i  j_z(\tau ,\xi)\,,\]
where the functions $j_\rho (\tau ,\xi)$ and  $j_z(\tau ,\xi)$ have been defined in accordance with expansion (\ref{j-two}).
Equation (\ref{lin_eq_1}) leads to the following equation for the function ${\mathfrak J}(\tau ,\xi)$:
\begin{equation}
        \label{lin_eq_com}
\partial_\tau {\mathfrak J} ~=~  - i \beta_1\,\left[ \partial_\xi^{\,2} {\mathfrak J}    + \frac{1}{2}\Bigl({\mathfrak J}   - \overline{\,\mathfrak J\,}\, \Bigr)\right] - \omega  \,.
\end{equation}

This equation has the following solution:

\begin{equation}
        \label{sol_1}
				{\mathfrak J}(\tau ,\xi) ~=~ \sum_{n} {\mathfrak j}_{\,n}\, e^{\,i\,[\,n\xi  +  n \sqrt{n^2 - 1}\,{\beta_1}\tau\,]}  ~-~ \omega\tau\,,   
\end{equation}			

where the complex-valued numbers ${\mathfrak j}_{\,n} \equiv const$.  The coefficients $ \overline{\,	{\mathfrak j}\,}_{\,-n}$  and  ${\mathfrak j}_{\,n}$ are related to each other as follows:

\begin{equation}
        \label{jn_jn-conj}
			\overline{\,	{\mathfrak j}\,}_{\,-n}     ~=~    2 \left[n\sqrt{n^2 -1} - n^2 + \frac{1}{2} \right]  {\mathfrak j}_{\,n} \,.
	\end{equation}			
As it was proved in the work  \cite{Tal},   condition  (\ref{constr_j_0}) will be  fulfilled.

{ Let $k_n = n/R$ means one-dimensional wave number, values $t$ and $S$ the real time and coordinate along the vortex ring. Taking into account definitions (\ref{tau_s}), the phase function $\Phi$  in Eq.(\ref{sol_1}) is written as follows:
\[\Phi(S,t) = k_n S  + \frac{\beta_1\Gamma}{4\pi R}k_n\sqrt{R^2k_n^2 -1}\, t\,.\] 
Therefore, the dispersion relation for our theory takes form
\begin{equation}
        \label{disp_rel}
  \omega = \omega(k_n) \equiv \frac{\beta_1\Gamma}{4\pi R}k_n\sqrt{R^2k_n^2 -1}\,,\qquad n=2,3,\dots \,.
\end{equation}

 This formula demonstrates that $\omega \propto k_n^2$ for large numbers $n$. This result corresponds to the dependency $\omega =  c k_n^2$  that was given in the paper \cite{BSS}. }

It is clear  that   solution  (\ref{sol_1}) is stable for $\tau \to \infty$.  	
Will the solution still be stable when taking into account the terms of the order $\varepsilon^k$, where $k = 2,3,\dots$? 
The problem of the stability of vortex ring perturbations has been investigated in the literature (see, for example, \cite{Protas}). This question lies outside of the scope of this paper, so we have no purpose to review the literature on this topic.
 Apparently, the answer is negative for the general case.   For what reason can we limit ourselves to terms of the order $\varepsilon$ in the case under consideration?
Our final purpose is to consider a system of interacting vortices. That is why, 
we intend to consider  vortex rings that exist during a short period of time  $t^v$ only.
So, for example, in a turbulent flow  \cite{Tenn}
\[  {t^v} ~\simeq~ \frac{2\pi}{|\boldsymbol{w}|}\,,\]
where the vector  $\boldsymbol{w}$  means  the vorticity.

As an intermediate final, we have the following set of variables which describes the dynamics of the perturbation amplitude  ${\mathfrak J}(\tau ,\xi)$:

		\begin{enumerate}
		\item The real variable ${\rm j}_{\,0}(\tau) = {\mathfrak j\,}_{\,0}  - \omega\tau\,$ that depends on ''time'' $\tau$ lineally;
\item The time-independent complex-valued  variable ${\rm j}_{\,-1} = {\mathfrak j\,}_{\,-1}$;
\item  The oscillating variables ${\rm j}_{\,-n}(\tau) = {\mathfrak j\,}_{\,-n}\,e^{i[\,n \sqrt{n^2 - 1}\,\beta_1\tau\,]}$ ($n = 2,3,\dots$) that generate helical perturbations of the original vortex ring.
\end{enumerate}

Let us define the value $\Delta R$:
\begin{equation}
        \label{Delta_R}
	\Delta R ~=~ \sqrt{R^2 - R_f^2}\,,
				\end{equation}
			where the constant $R_f$ was introduced by the inequality (\ref{R_f}).	
To describe the excitations of  vortex ring as a whole  (not only the tangent vector ${\boldsymbol j}$), we must add the value  $\Delta R$, coordinates ${\boldsymbol q}$ and the components of the unit vector ${\boldsymbol{e}_z}$ as dynamical variables.
Because the vortex in question is  oriented in space  arbitrarily,  vector ${\boldsymbol{e}_z}$ has two independent components. 

 Also we include the circulation $\Gamma$ as an additional dynamical variable. Such a step, which is characteristic of the proposed approach, makes it possible to take into account the movement of the surrounding fluid  in a minimal way.  From the viewpoint   of further theory construction, this step will allow   to define extended set of the variables  that characterize the vortex dynamics.
As a consequence, we apply the  many-body   interaction  theory  to describe the interaction of vortices.

\vspace{3mm}

\section{New set of the dynamical variables}

~~~ Before we quantize the considered dynamical system, let's transform the set of variables introduced above.
Fist of all, we postulate here the well-known formula \cite{Batche} for the canonical momentum ${\boldsymbol p}$:
	  
   \begin{equation}
        \label{p_and_m_st}
        {\boldsymbol p} ~=~ \frac{\varrho_0}{2 }\,\int\,\boldsymbol{r}\times\boldsymbol{w}(\boldsymbol{r})\,dV\,.
        \end{equation}

The vector  ${\boldsymbol{w}}(\boldsymbol{r})$  means  the vorticity and the constant 
$\varrho_0$ means the  fluid density.                
      As well-known,   the vorticity of the  closed vortex filament is calculated as follows
   \begin{equation}
        \label{vort_w}
     {\boldsymbol{w}}(\boldsymbol{r}) ~=~  \Gamma
                  \int\limits_{0}^{2\pi}\,\hat\delta(\boldsymbol{r} - \boldsymbol{r}(\xi))\partial_\xi{\boldsymbol{r}}(\xi)d\xi\,,
       \end{equation}            
   where the symbol $\Gamma$ denotes the circulation  and the symbol  $\hat\delta(\xi)$ means $2\pi$-periodical $3D$  $\delta$-function.
		Taking into account the formulae  (\ref{proj_r}),   (\ref{constr_j_0}) and   (\ref{vort_w}),    we deduce the  expression for the canonical momentum in form
	\begin{equation}
        \label{impuls_def}
    {\boldsymbol{p}}  ~=~    \varrho_0 {R}^2 \Gamma             {\boldsymbol f} \,,            
      \end{equation} 

where
\[  {\boldsymbol f} ~=~ \frac{1}{2}\int\limits_{0}^{2\pi} \int\limits_{0}^{2\pi} \left[\, {\xi - \eta}\,\right]\,{\boldsymbol j}(\eta)\times{\boldsymbol j}(\xi)d\xi  d\eta\,.    \]

Next, we introduce the variables $p_x$,  $p_y$ and $p_z$  instead of the variables $\Gamma$ and two independent components of the vector   ${\boldsymbol{e}_z}$.
However, this redefinition of the set  
 $\{\Gamma, {\boldsymbol{e}_z}, \Delta R, {\boldsymbol j}_p(\xi)\}$ leads to the problem: variations of the variables ${\boldsymbol{p}}$ and ${\boldsymbol j}(\xi)$ are  not independent in general due to equality  (\ref{impuls_def}).
Let us introduce   the set
\begin{equation}
\label{new_set}
{\cal A}^{\,\prime} ~=~ \bigl\{\, {\boldsymbol p}\,,  {\boldsymbol{q}}\,; \Delta  R\,, {\boldsymbol j}_p(\xi) \,\bigr\}\,,
 \end{equation}
where the variable  $\boldsymbol{q}$ was  defined by the formula  (\ref{q_def}).
Can we consider the set  ${\cal A}^{\,\prime}$ 
as the set of fundamental independent variables of the model?
To answer this question, we investigate the variation $\delta{\boldsymbol p}$ that is generated by the variation of the function ${\boldsymbol j}_0(\xi)$. 
As it has been proved in the paper \cite{Tal}, this variation   of the initial momentum ${\boldsymbol p}$ is transverse  and can be represented in a complex form as
\begin{equation}
\label{var_p}
\delta{\boldsymbol p} ~=~\delta p_x + i \delta p_y ~=~  2\pi\varrho_0 {R}^{\,2} \Gamma\, {\mathfrak j}_{-1}\,.
\end{equation}

Therefore, the constraint
\begin{equation}
\label{con_j-1}
{\mathfrak j}_{-1} ~=~ 0
\end{equation}
makes the variations $\delta{\boldsymbol p}$ and $\delta{\boldsymbol j}(\xi)$ independent.

In order to understand what constraint (\ref{con_j-1}) means, we will consider how the excitation of any mode 
${\mathfrak j\,}_{\,-n}$,  $n = 0, 1, 2, \dots$   transforms the original vortex ring
(\ref{proj_r}) -- (\ref{tang_v}).

{\bf 1.} The case ${\mathfrak j\,}_{\,-n} = 0$  ~($n = 1, 2, \dots$),    ~${\mathfrak j\,}_{\,0} \not= 0$.

\begin{figure}[t]
\includegraphics[width=4.5in]{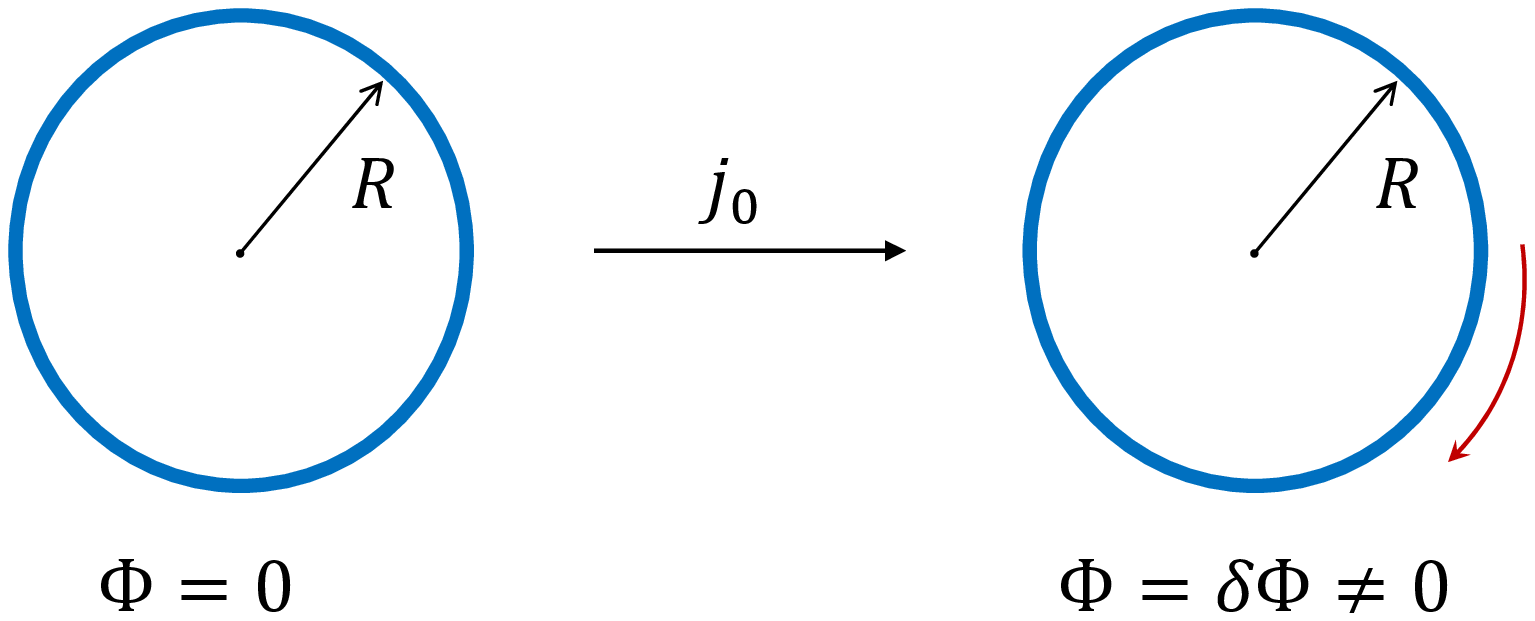}
\vspace{-15mm}
\caption{Vortex ring perturbation generated by the variable ${\mathfrak j}_{\,0}$}
\label{Pic_svt1}
\end{figure}

This excitation of the tangent vector ${\boldsymbol j}_0(\xi)$ transforms  vector ${\boldsymbol j}_0(\xi)$ and corresponding vector ${\boldsymbol r}_0(\xi) = R\, {\mathfrak r}_0(\xi) $  (see (\ref{proj_r}))   as follows:
\begin{eqnarray}
\label{j_trn0}
{\boldsymbol j}_0(\xi) & \longrightarrow & {\boldsymbol j}_0(\xi) + \varepsilon\, {\mathfrak j}_{\,0}(\tau){\boldsymbol e}_\rho(\xi)\,,\\[2mm]
\label{r_trn0}
{\boldsymbol r}_0(\xi) & \longrightarrow & {\boldsymbol r}_0(\xi) - \varepsilon\, R\,{\mathfrak j}_{\,0}(\tau){\boldsymbol e}_\xi(\xi)\,.
\end{eqnarray}
Thus, the considered excitation generates ''small rotation'' of the initial circular vortex ring.
This rotation models the appearance of a non-zero flow $\delta\Phi$  in the vortex core.
Indeed, the corresponding  ''master'' equation   (\ref{LIE_pert}) has the exact solution

\begin{equation}
        \label{our_sol}
 {\mathfrak r}(\tau ,\xi) ~=~ \Bigl(\, \frac{q_x}{R} +  \cos(\xi +\phi_0 +\varepsilon\,\omega\tau)\,, \frac{q_y}{R} +  \sin(\xi +\phi_0 + \varepsilon\,\omega\tau  )\,, \frac{q_z}{R} + \beta_1 \tau \,\Bigr)\,, 
\end{equation}

where the constant   $\phi_0 = -\varepsilon\,{\mathfrak j}_{\,0}(0)$. The case $\varepsilon << 1$ gives the perturbations (\ref{r_trn0}).  Schematically, this case is shown  on Fig.\,\ref{Pic_svt1}.

{\bf 2.} The case ${\mathfrak j\,}_{\,-n} = 0$  ~($n = 2, 3, \dots$),    ~${\mathfrak j}_{\,-1} \not= 0$. As regards the mode ${\mathfrak j}_{\,0}$, it is non-zero due to the linear dependence on ''time'' $\tau$. Correspondent excitations were investigated in the previous item so  it will be ignored here.

Let's denote $\varphi = |\,{\mathfrak j}_{\,-1}|$ and $\gamma = \arg{\mathfrak j}_{\,-1}$.
We can assume that $\gamma = 0$   if we  change the initial point for the angle  $\xi$. In accordance with   our assumptions, the considered excitation transforms the tangent vector
  ${\boldsymbol j}_0(\xi)$ as follows:
	
\begin{eqnarray}
{\boldsymbol j}_0(\xi) & = & {\boldsymbol e}_\xi(\xi) ~\longrightarrow~
{\boldsymbol e}_\xi(\xi) + \varepsilon\varphi\cos\xi\,{\boldsymbol e}_\rho(\xi) - 
\varepsilon\varphi\sin\xi\,{\boldsymbol e}_z ~ = \nonumber \\[2mm]
~& = & \Bigl(- \sin\xi +  \varepsilon\varphi\cos^2\xi\Bigr)\,{\boldsymbol e}_x ~+~ 
\Bigl(\cos\xi + \frac{\varepsilon\varphi}{2}\sin 2\xi\Bigr)\,{\boldsymbol e}_y 
~+~\nonumber \\[2mm]
~&+& \varepsilon\varphi\sin\xi\,{\boldsymbol e}_z ~\simeq~
 -~ \sin\xi\,\Bigl[{\boldsymbol e}_x  - \varepsilon\varphi\,{\boldsymbol e}_z\,\Bigr] ~+~ \cos\xi\,{\boldsymbol e}_y\,.
\label{j_trn1}
\end{eqnarray}

\begin{figure}[t]
\includegraphics[width=4.5in]{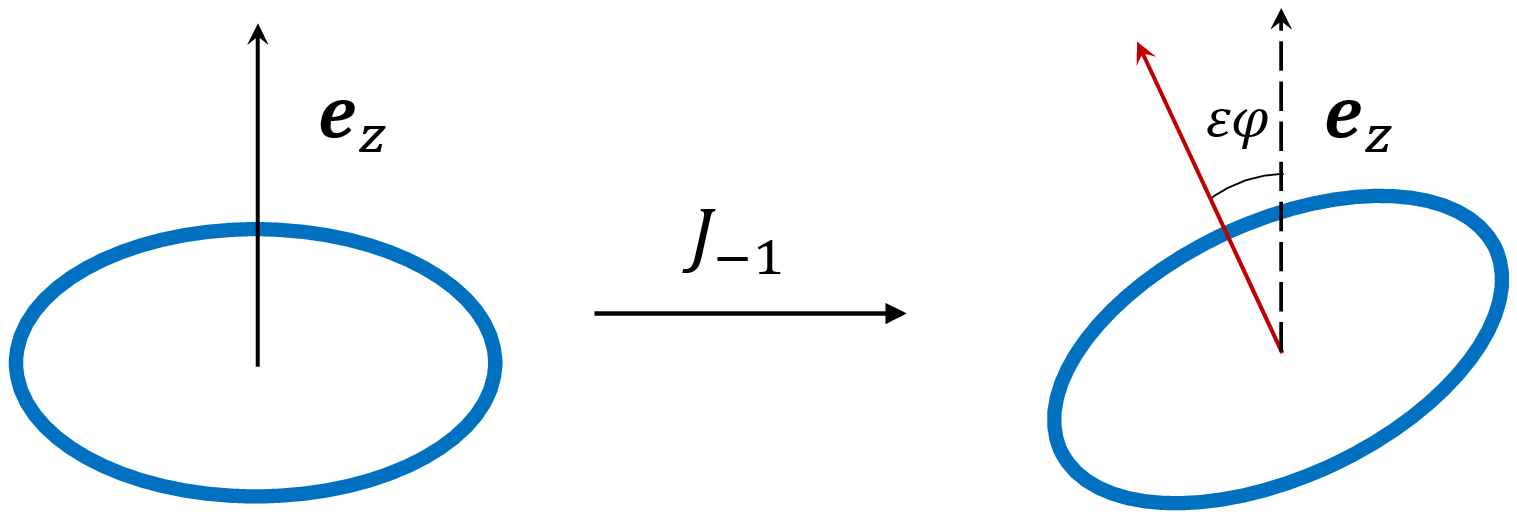}
\vspace{-15mm}
\caption{Vortex ring perturbation generated by the variable ${\mathfrak j}_{-1}$}
\label{Pic_svt2}
\end{figure}

Thus, this perturbation corresponds to a rotation around the vector ${\boldsymbol e}_y$ by a ''small angle'' $\varepsilon\varphi$.
This case is demonstrated on Fig.\,\ref{Pic_svt2}.
The direction of   the unit vector  ${\boldsymbol e}_y$ in the  
plane $X0Y$  was fixed above by the parameter $\gamma$.
We consider SO(3) - invariant theory. Because the perturbation (\ref{j_trn1}) can be compensated by the space rotation, we assume the constraint (\ref{con_j-1}) is fulfilled.

\begin{figure}[th]
\includegraphics[width=4.5in]{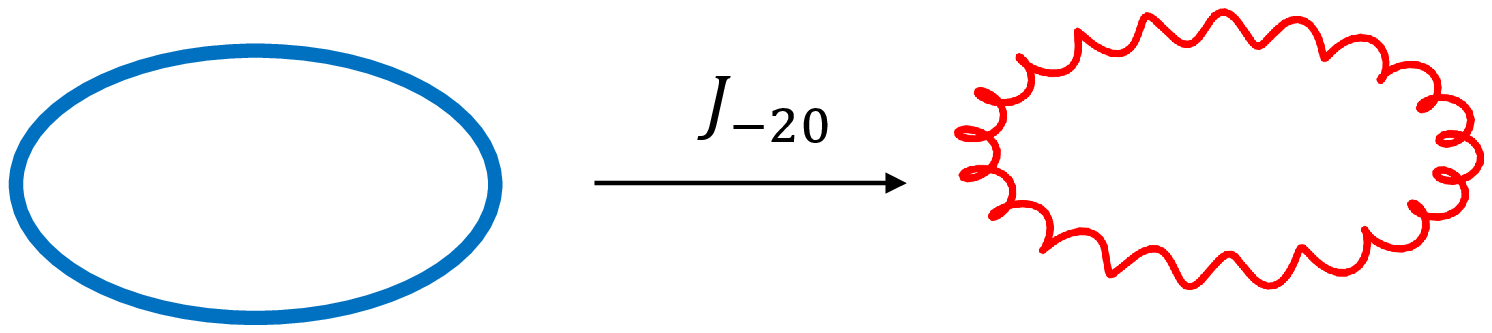}
\vspace{-15mm}
\caption{Vortex ring perturbation generated by the variable ${\mathfrak j}_{-n}$, ~$n\not=  0,1$.}
\label{Pic_svt3}
\end{figure}

{\bf 3.} The case ${\mathfrak j\,}_{\,-m} \not= 0$,  ~${\mathfrak j\,}_{\,-n} = 0$ 
 ~($n = 2, 3, \dots, m-1, m+1, \dots$).
Those perturbations lead to the helical deformations of the initial circular vortex ring.
We demonstrate this type of the perturbation for $n = 20$ schematically on Fig.\,\ref{Pic_svt3}

 The formula  (\ref{impuls_def}) leads to equality
\begin{equation}
\label{p_Gamma}
{\boldsymbol p}  =  \pi\varrho_0 {R}^{\,2} \Gamma\,{\boldsymbol e}_z \,.
\end{equation}

This equality demonstrates the independence of the variables ${\boldsymbol p}$ and
${\boldsymbol j}_p(\xi)$.   

 The circulation $\Gamma$ becomes the function on the set of new variables.
Indeed, due to the equality  (\ref{p_Gamma})
the following equation  is fulfilled: 
\[{\boldsymbol p}^2   =  \pi^2\varrho_0^2 {R}^{\,4} \Gamma^{\,2}\,.\]
For subsequent purposes, we define the  variables $\varpi$ and
 $\chi$ instead of variables $\Delta R$ and ${\mathfrak j\,}_{\,0}(\tau)$. 
The relevant definitions are as follows:

\[  \varpi = \frac{\Delta R}{R_f}\cos({\mathfrak j\,}_{\,0} +\omega\tau)\,, \qquad   \chi = \frac{\Delta R}{R_f}\sin({\mathfrak j\,}_{\,0} +\omega\tau)\,.\]
It is clear that these variables are similar to the variables of a classical harmonic oscillator.
Finally, we  postulate the set
\begin{equation}
\label{new_set1}
{\cal A}^{\,\prime} ~=~ \bigl\{\, {\boldsymbol p}\,,  {\boldsymbol{q}}\,;\quad \varpi\,, \chi\,;\quad {\mathfrak j\,}_{\,-n}\,, \, \overline{\mathfrak j\,}_{\,-n}\, \quad (n ~=~ 2,3,\dots)  \,\bigr\}
 \end{equation}
as the set of fundamental independent variables for the considered theory.
Therefore, 
 the circulation $\Gamma$ becomes  the  function  of the   variables (\ref{new_set1}). This function is defined from the equation
\begin{equation}
\label{main_con}
{\boldsymbol p}^{\,2}   ~=~  \pi^2\varrho_0^2 {R_f}^{\,4}\Bigl(1 + \varpi^{\,2} + \chi^{\,2} \Bigr)^{2}\, \Gamma^{\,2}\,.
\end{equation}

\vspace{3mm}

\section{The vortex quantization \\  and the  spatio - temporal scale}

~~~In this section, we  construct  quantum theory of the closed vortex filament in terms of the variables 
(\ref{new_set1}). 
This  theory has  been 
investigated in the paper \cite{Tal} 
for the  special  case $\omega = 0$  (see Eq.\,(\ref{LIE_pert})), $R = const$,  but without the restriction ${\mathfrak j}_{\,-1} = 0$. 
The case $\varepsilon\omega = \beta_2$ (where the number  $\beta_2$ takes  arbitrary values) has been considered in the paper \cite{Tal_PhRF} for the circular loops only.
Of course, in order to quantize the theory in a canonical way, we need a well-defined Hamiltonian structure.  This issue is discussed  in the cited authors'  works   in detail.
The theory being developed in this study includes elements of both cases considered.
Therefore, we will go through  the main points of the quantization scheme without going into details.

The  introduced set (\ref{new_set1}) of classical dynamic variables allows us to specify the
canonical  quantization scheme.  Indeed, it is natural to assume that  the quantum states of the considered vortex loop are the vectors of the Hilbert space
\begin{equation}
	\label{space_quant}
	\boldsymbol{H}_1  ~=~  \boldsymbol{H}_{pq} \otimes   \boldsymbol{H}_b \otimes   \boldsymbol{H}_j \,,
	\end{equation}
			where the symbol   $\boldsymbol{H}_{pq}$  denotes the Hilbert space  of a free structureless particle in $3D$ space ${\sf R}_3$.  In this paper we assume $\boldsymbol{H}_{pq} = L^2({\sf R}_3)$.  	The	symbol $\boldsymbol{H}_b $ denotes		the Hilbert space  of the quantized harmonic oscillator with classical variables $\chi$ and $\varpi$ (see definitions of these quantities before the formula (\ref{new_set1})).
		In terms of the creation and annihilation operators  $\hat{b}^+$ and $\hat{b}$,  this space is formed by the vectors
		\[|\,s\rangle   ~=~  \frac{1}{\sqrt{s!}} (\hat{b}^+)^s    |\,0_b\rangle   \qquad 
		[\,\hat{b}, \hat{b}^+] ~=~ \hat{I}_b\,, \quad \hat{b}|\,0_b\rangle ~=~ 0\,,  \]
			where $|\,0_b\rangle \in \boldsymbol{H}_b$. 	
								The	symbol $\boldsymbol{H}_j $ denotes the space for the  number of the harmonic oscillators    which correspond to the variables ${\mathfrak j}_{\,-n}$.		 
	To simplify further consideration, we assume that   a finite number  $M$ of modes 
	 ${\mathfrak j}_{\,-n}$ are excited only. 
	Thus, 	the following  conditions are fulfilled for some number $M$:
				\[{\mathfrak j}_{\,-n} ~=~ 0\,, \qquad \forall\,n > M\,.\]				
							
				The creation and annihilation operators which are defined in the space 	$\boldsymbol{H}_j $, have standard commutation relations
				\[[\,\hat{a}_m, \hat{a}_n^+] = \delta_{mn}\hat{I}_F\,, \quad \hat{a}_m|\,0\rangle = 0  \, \quad  (m,n = 2,3,\dots )\,,\qquad |\,0\rangle \in    \boldsymbol{H}_j\,.\]  				
			 
On this ground, we quantize our (internal) variables  $\varpi$, $\chi$  and ${\mathfrak j}_{\,-n}$ in this way:
\[ \chi ~\to~ \sigma_{ph}\, \frac{\hat{b} + \hat{b}^+}{\sqrt{2}}\,, \qquad 
\varpi ~\to~ \sigma_{ph}\, \frac{\hat{b} - \hat{b}^+}{i\sqrt{2}}\,,
\qquad 
{\mathfrak j}_{\,-n} ~\to~ \sigma_{ph}\, \hat{a}_n\,, \]
where $n = 2,3, \dots$ and the dimensionless constant
\[\sigma_{ph} ~=~ \sqrt{\frac{\hbar}{t_0{\mathcal E}_0}} ~=~ \sqrt{\frac{\hbar}{\mu_0 v_0 R_0}}\,\]
depends on the Planck constant $\hbar$. 
As regards of the  constants $t_0$  and  ${\mathcal E}_0$, they define the  time and energy scales in constructed classical  theory.  In our model
\[  t_0 = \frac{R_f}{v_0}\,, \qquad    {\mathcal E}_0 = \mu_0 v_0^2\,,\]
where the constant $v_0$ is the speed of sound in a fluid.
  The mass parameter $\mu_0$ is a central charge for the central extended Galilei group.  The relevance  of this group for our model  was discussed in the cited author's works in detail.  We  use the variable parameter $\mu_0$ together with  ''natural'' mass parameter $\tilde\mu_0 = \rho_0 R_f^3$.
Possible values of the parameter $\sigma_{ph}$ were discussed in  the paper \cite{Tal_PhRF}. For example,  an estimated value of this parameter\footnote{Instead the constant $R_f$, the ''pipe radius'' $R_0$ was  used to define the value $t_0$ in the paper \cite{Tal_PhRF}. }   was deduced   as  $\sigma_{ph} \simeq 10^{-3}$.
Apparently, other parameter values can also be used: we will not discuss this issue in this paper.
It is appropriate to note  that proposed quantization scheme is ''inverse realization''  of Lord Kelvin's old idea that any particles can be considered as some kind of vortex-like structures  \cite{Thom}. In our approach, we describe the closed vortex filament  as some structured particle.

After quantization of the equation (\ref{main_con}), we have the following spectral problem
for possible values of circulation $\Gamma$:
\begin{equation}
        \label{for_Gamma1}
\left[\hbar^2\Delta  + \pi^2 \varrho_0^2 \Gamma^2 R_f^4  \Bigl(1  +  \sigma_{ph}^2\,
(\hat{b}^+ \hat{b} + 1/2)  \Bigr)^2\right]|\Psi\rangle ~=~0\,, 
	\end{equation}
where $|\Psi\rangle \in  \boldsymbol{H}_1$. 
{ This equation is the eigenvalue problem for the circulation only. The energy spectrum  as well as the  quantum  energy  states can be deduced after quantization of classical energy (see Eq. (\ref{E_general})) and subsequent solving corresponded spectral problem. These issues   
 lie outside of the item of this study. 
Some specific moments of  this problem  have been explored in other works of the author: see, for example, a recent paper 
\cite{Tal_24}.}

Let us consider the following example. We assume that the fluid moves  in the domain $V$, which is determined by the following boundary conditions:
\begin{equation}
\label{domain_V}
 x^2 + y^2  \le R_0^{\,2}\,, \qquad    z \in [0, 2\pi R_1]\,\, (mod\,2\pi R_1)\,, \quad R_1 ~=~ const\,. 
\end{equation}
This domain  describes  a   round   tube with a radius $R_0$  in the shape of a torus of radius $R_1$. As it seems, the domain models a long  thin pipe  in case $R_1 >> R_0$.  We consider this condition to be fulfilled further.
The eigenvalues $-\lambda^2$  of the Laplace operator $\Delta$ in the  domain $V$  take the following values:
\[  \lambda_{m,\ell,k}^2  ~=~  \left(\frac{m}{2 R_1}\right)^2 +  \left(\frac{\zeta^{(\ell)}_k}{R_0}\right)^2\,, \qquad  \ell ~=~ 0,1,2,\dots\,\quad    m, k ~=~ 1,2,\dots\,,\]
 where the  quantities $\zeta^{(\ell)}_k$, $k = 1,2,\dots$  stand for  zeros of the Bessel function $J_\ell(\rho)$.
Therefore,
\begin{equation}
\label{circ_t}
				\Gamma_{s;\,m,\ell,k}  ~=~  \pm\,\frac{\hbar R_f\lambda_{m,\ell,k} }{~\tilde\mu_0 \bigl( 1 +  \sigma_{ph}^2(s + 1/2)\bigr)  ~}\,,
\end{equation}
where natural numbers $s, m,\ell = 0,1,2,\dots\,$  and number  $k = 1,2,\dots$.

This result differs significantly from that in the Gross-Pitaevsky model, where the quantized circulation is proportional to an integer: $\Gamma_n\propto\hbar n$. 
Such a formula for $\Gamma_n$  has been considered from different points of view (see recent paper \cite{BelRic}, for example).
However,  quantum hydrodynamics and the theory of super-fluidity are not the same things
\cite{Feyn_1}.  As the author believes, this statement opens up possibilities for building theories of quantum fluids outside the approach of the Gross-Pitaevsky model.

The paper \cite{Tal_PhRF} provides the arguments in favor of generalizing of the standard formula   for the circulation. Briefly, these arguments are as follows.

\begin{itemize}
\item The known experimental measurements of the magnitude of $\Gamma$ were made for special single vortices in a superfluid, and not in any complex many-vortices flow.  Therefore, it cannot be excluded that the  conventional  formula  $\Gamma_n = n \hbar/\mu$ 	   { can be generalized }  for    complicated vortex systems in a some way. 
\item The rule $\Gamma_n = n \hbar/\mu$ is usually postulated.  
 As   has been repeatedly stated in the literature, such quantization rules are similar to the quantization rules in the old Bohr quantum theory.
The author believes that quantum  values $\Gamma_n$  should be deduced  from the general postulates of quantum theory, and not postulated separately.
\item The argument exists 
  that  standard rule $\Gamma_n = n \hbar/\mu$ is the consequence of an unambiguity of the wave function $\Psi$ for certain
 quasi-particles\footnote{In a two-fluid model, which we are not considering here.}, because $\arg\Psi \propto  \Gamma_n$.
Here the author would like to note the following. 
We consider the       fluid  environment  where the closed vortex loops with  very small  thickness are present.   This environment  can be considered as a
 realization  of multi-connected space.  Thus, any quasi-particle here can possess the fractional statistics. 
In this case, the wave function of a quasi-particle can get a phase multiplier when moving along a closed path around a vortex filament.
\end{itemize}

Now  we discuss the asymptotic behavior of the values $\Gamma_{s,m,\ell,k}$.
\begin{enumerate}
\item The number $s$ is small, the number ${\sf i}\to\infty$, where the   number ${\sf i}$ denotes the number $m$, or number $\ell$, or number $k$. All other numbers are fixed.   As it is well known, the asymptotic behavior of values  $\zeta^{(\ell)}_k$  for large values
$\ell$ and $k$ will be   $\zeta^{(\ell)}_k  \simeq (3\pi/4) + (\pi/2)\ell + \pi k$.
The asymptotic expression of the value  $\lambda_{m,\ell,k}$ for large number $m$    is also proportional to $m$.
Therefore, we obtain the following asymptotic behavior for large number ${\sf i}$:
\begin{equation}
\label{conv_circ}
\Gamma_{\sf i} ~\longrightarrow~ \frac{\hbar\, {\sf i}}{\mu_1} ~+~ {\mathcal O}(\hbar^2) \,,
\end{equation}
where the constant $\mu_1$ has the dimension of mass. 
Let's note that formula (\ref{conv_circ}) contains the anomalous terms that are proportional to the constants $\hbar^n$, where $n = 2,3, \dots$,  in comparison with the conventional formula  $\Gamma_{\sf i} =  ({\hbar }/{\mu_1}){\sf i}$.  
\item The number $s$ is large:
\[ s ~>>~ \frac{1}{\sigma_{ph}^2}\,.\] 
All other numbers are fixed.
In this case\footnote{In a sence, this limit ($s \to \infty$) is equivalent to the case $R_f=0$ in the condition (\ref{R_f}). Such a case was considered in the paper  \cite{Tal_PhRF} with  corrections of some definitions. }, the quasi-classical behavior of the value $\Gamma$ is observed. Indeed, because
$\sigma_{ph}^2 \propto \hbar$,
\[\Gamma_s ~\longrightarrow~ \frac{t_0{\mathcal E}_0}{~\tilde\mu_0 (s + 1/2)} \, R_f\lambda_{m,\ell,k}\,.\]
 In general,  the set $\{\Gamma_s\}$   has  same fractal structure as the sequence $1/s$ in a neighborhood of the point   $(1/s) \to 0 $.  
\end{enumerate}

Of course, in real systems, the quantum number $s$ cannot take infinite values.
Indeed,   in accordance with definition of the variables $\chi$,  $\varpi$ and definition (\ref{Delta_R}), the vortex radius $R$ is quantized as follows:
\[ R^{\,2} ~\longrightarrow~ \hat{R}^{\,2} ~=~ R_f^{\,2}
\left[1 + \sigma_{ph}^2\left({\hat b}^+ {\hat b} + \frac{1}{2}\right)\right]\,.\]   
Thus, the eigenvalues  ${R}_s$ take following values:
\[ {R}_n ~=~ R_f\sqrt{1 + \sigma_{ph}^2\left(s + \frac{1}{2}\right)} \,,
\qquad  s ~=~ 0,1,\dots, N\,.\]
The finite number $N$ appears here because the value ${R}_s$ is restricted by the    size of the domain $V$:
 the inequality $R_s \le R_0$ is fulfilled  due to  the obvious physical and geometrical   reasons. This inequality leads to the
limitations for circulation values  $\Gamma$
 in the considered theory:
\begin{equation}
\label{Gamma_min}
   \Gamma ~\ge~  \Gamma_{min} ~=~  \frac{\hbar \zeta^{(0)}_1}{\varrho_0 R_0^3}\,.
	\end{equation}
	This conclusion makes it possible to clarify case $s >> {1}/\sigma_{ph}^2$, which was discussed above. In this case, the set $\{\Gamma_s\}$ has not a fractal, but a quasi-fractal structure.

	Let's explore the question of what is the minimal distance $\Delta\Gamma_{min}$   between the circulation levels near the point
	$\Gamma_{min}$.  
For the assumption that $s \simeq N$ ($s < N$), we deduce from the formula	(\ref{circ_t}):
\begin{eqnarray}
\Delta\Gamma_{min} & = & \min_{m,\ell,k} \Big\vert\Gamma_{s;\,m,\ell,k} - \Gamma_{s+1;\,m,\ell,k}\Big\vert ~= \nonumber\\[2mm]
\label{quant_min}
~  & = & \frac{\hbar}{\tilde\mu_0}\sigma_{ph}^2\zeta^{(0)}_1\left(\frac{R_f}{R_0} \right)^{\!5}
\left[1 +{\mathcal O}\Bigl(\frac{1}{s}\Bigr) \right]\,.
\end{eqnarray}
We have deduced a non-trivial and some unexpected  result that  $\Delta\Gamma_{min} \propto \hbar^2$, because 
$  \sigma^2_{ph}  \propto \hbar $.
Obviously, $\Delta\Gamma_{min}<< \hbar/\mu_H$, where the constant $\mu_H$ means the atomic mass in  some quantum fluids (for example, the mass of the ${\,}^4\!H\!e$ atom).
Practically, the spectrum of the value $\Gamma_{s;\,m,\ell,k}$    is quasi-continuous  near the point $\Gamma_{min}$.

The quantum states that correspond to the circulation $\Gamma \approx \Gamma_{min}$ can be realized for the ''very large'' vortex loops (with radius $R_s \simeq R_0$). As regards of the ''micro-vortices'' with radius $R_s \simeq R_f$ (small values of the quantum  number $s$), the formula (\ref{conv_circ}) demonstrates that circulation $\Gamma$  tends to the conventional value\footnote{We can choose the parameters of our theory so that $\mu_1 = \mu_H$ for example.}.
In this regard, it is interesting to consider the structure of the flow in some  domain $V_1 \subset V$.  
In such a domain, we can observe both ''small'' vortex loops contained entirely in the domain $V_1$, and vortex filaments extending beyond its boundaries.  These filaments are the parts of  ''large'' loops in the domain $V$.

So far, we have considered the evolution of a vortex loop in some ''conditional'' time, which can be defined as
\[ \tilde t ~=~ t_0 \tau\,.\]
Next, we will consider a quantum vortex loop ${\boldsymbol r}(t,s)$ in real time $t$.
Let's assume that our system is in a pure quantum state
  $|s,m,\ell,k\rangle$, which is characterized by quantum numbers $s,m,\ell,k$.
In accordance with formulas (\ref{tau_s}) and (\ref{proj_r}),  this loop is described as follows:
\begin{equation}
\label{r_scales}
{\boldsymbol r}(t, S) ~=~  R_s \,{\mathfrak r}\bigl(\,t/{\mathfrak T}_{[w]}\,,\,S/2\pi R_s\,\bigr)\,,
 \end{equation}
where scale factor  ${\mathfrak T}_{[w]}$ is defined as
\[  {\mathfrak T}_{[w]} ~=~   \frac{4\pi R_s^{\,2}}{\Gamma_{s,m,\ell,k}}\,,\qquad  [w] = \{s,m,\ell,k\}\,.\]
Therefore, every pure quantum state $|s,m,\ell,k\rangle$  corresponds to the evolution of the vortex on its own time scale.  In accordance with the results obtained above
\[  {\mathfrak T}_{[w]} ~\in~  (\,0\,,\,{\mathfrak T}_{max}\,)\,,\]
where the constant ${\mathfrak T}_{max}$ depends on the size and form of the domain $V$. For example, the value of the constant ${\mathfrak T}_{max}$ in our case is obtained as follows: 
\[ {\mathfrak T}_{max} ~\simeq~ \frac{4\pi \varrho_0 R_0^{\,5}}{\hbar\, \zeta^{(0)}_1}\,.\]
It is appropriate to note here that the set $\{{\mathfrak T}_{[w]}\} $ has a fractal structure for the large numbers $m,\ell,k$.
Taking into account formula (\ref{our_sol}), we find the absolute value of the velocity of the vortex loop in the flow:
\[  {\mathfrak v}_{[w]} ~=~  \beta_1\frac{R_s}{{\mathfrak T}_{[w]}}\,,\qquad  [w] = \{s,m,\ell,k\}\,.\]

{ Regardind the dispersion relation (\ref{disp_rel}) for the Kelvin waves of the vortex ring, we have the  hierarchy of these relations   after quantization:
\[\omega = \omega_{[w]}(k_n) \equiv \frac{\beta_1\Gamma_{s,m,\ell,k}}{4\pi R_s}k_n\sqrt{R_s^2k_n^2 -1}\,,\qquad n=2,3,\dots \,.\]}

\vspace{3mm}

\section{The creation and annihilation \\ of the quantized vortices}

~~~ Of course, the evolution of vortex loops in a developed turbulent flow does not occur in accordance with the LIE equation\footnote{The possibility of generalizing the equation to describe a developed turbulent flow is discussed by the author in the paper \cite{Tal_24}}. 
However, it is of interest to describe the  origin  of turbulence in  
a quantum fluid. This phenomenon may be a consequence of the fact that
  single isolated vortices  begin to appear and disappear.  
In this case,
we  assume that   individual vortices can be described by the equation  (\ref{LIE_pert}). Accordingly, we can apply the theory discussed above.

As it is mentioned at the end of paragraph 2, the    variables  (\ref{new_set1})   make it possible  to describe the interaction of vortices in  terms of
 the  many-body theory.
This issue has been discussed more detail in the paper \cite{Tal_PhRF}. Here  we  go through   the main points only.
\begin{itemize}
\item The space of a quantum states  is the Fock space 
\[{\mathfrak H}  ~=~  \bigoplus_{N=0}^{\infty}\boldsymbol{H}_N \,,  \] 
where
\[ \boldsymbol{H}_N ~=~   \underbrace{\,\boldsymbol{H}_1 \otimes  \dots \otimes \boldsymbol{H}_1\,}_{N}   
\,.\]
\item   Hamiltonian
\begin{equation}
\label{ham_q_full}
\hat{H} ~=~   \hat{H}_0 +   \hat{U}\,, \nonumber
\end{equation}
where operator $\hat{H}_0$ means the Hamiltonian for the system of the non - interacting vortices
\cite{Tal_PhRF}\footnote{As already mentioned at the beginning of this  study,  both the vortex energy and the single vortex  Hamiltonian  for the case ${\sf a} \to 0$ are defined by group-theoretical methods. The details were discussed in the cited works of the author. }.
 \item  The interaction Hamiltonian  $\hat{U}$ is written as follows:
\begin{equation}
\label{hat_U}
\hat{U} ~=~  \sum_{m,n = 1}^\infty \varepsilon_{mn} \hat{U}_{m\leftrightarrow n}\,.
\end{equation}
The sequence  $\varepsilon_{m n}$ is some finite decreasing  sequence.
The constants $\varepsilon_{m n}$ are  the coupling constants which define  the  intensity of the vortex  interaction.
Operators  $\hat{U}_{m\leftrightarrow n}$ define the  reconnection of the considered vortex filaments in the flow.  
Each such operator describes the transformation of  $m$ vortex rings into $n$  rings and vice versa, $n \to m$.  
\end{itemize}
Thus, the specific choice of the operator $\hat{U}$ corresponds to the fixation of a specific model of the system of interacting vortices. 
{ In genetal, the quantum theory of irreversible processes should lead to non-unitary evolution.  This hypothesis has been discussed in the book \cite{Prigogine} many years ago.  Therefore, the 
terms $\hat{U}_{m\leftrightarrow n}$ may be both hermitian and non-hermitian.  
For example, the theory with non-hermitian interaction   $U = \varepsilon\, \hat{U}_{1\to 2} $ was investigated in the paper  \cite{Tal_PoF23}, where the development of the turbulent flow in it's initial stage has been considered. }

It should be emphasized that the suggested approach to the vortices interaction is not completely reduced to the theory of many bodies.
Indeed,  the processes $n \leftrightarrow m$, where $n,m = 1,2,3,\dots$, and also  processes $n \leftrightarrow 0$, where $n \ge 2$ (''vacuum loops'')  are valid in the many - body theory.
As for the processes $1 \to 0$ or $0 \to 1$,    they are impossible in the non-relativistic  theory, because  massive bodies (particles) cannot disappear into "nothing" and cannot arise "out of nothing".


  This is not the case in our approach.   Indeed, variables ${\boldsymbol p}$ and  ${\boldsymbol q}$ describe the  virtual particles corresponding to the external degrees of freedom of the vortex ring.  Unlike  real  particles in a vacuum, single vortices can  appear and disappear  in the surrounding fluid for some physical reasons. 
Such reasons  may be either some fluctuations in the flow,  or processes in the boundary layer, etc.
{ For example, the process $1 \to 0$  may be caused by dissipative forces. Note that dissipative phenomena in superfluids continue to be investigated at the present time (see, for instance, \cite{Tang}).  As for our approach, dissipative forces can be included in the consideration if we add certain  non-hermitian terms to Eq.(\ref{hat_U}).  }
A specific conditions  that makes   enable  such processes in the theory being developed is the 
''very small'' values of the quantities $\Gamma_{min}$ and   $\Delta\Gamma_{min}$ (see Eqs. (\ref{Gamma_min}) and (\ref{quant_min}) ).
{ Therefore, we can assume that the vortex creation        are permissible and can lead to subsequent development of turbulence.}

\begin{figure}[th]
\includegraphics[width=4.5in]{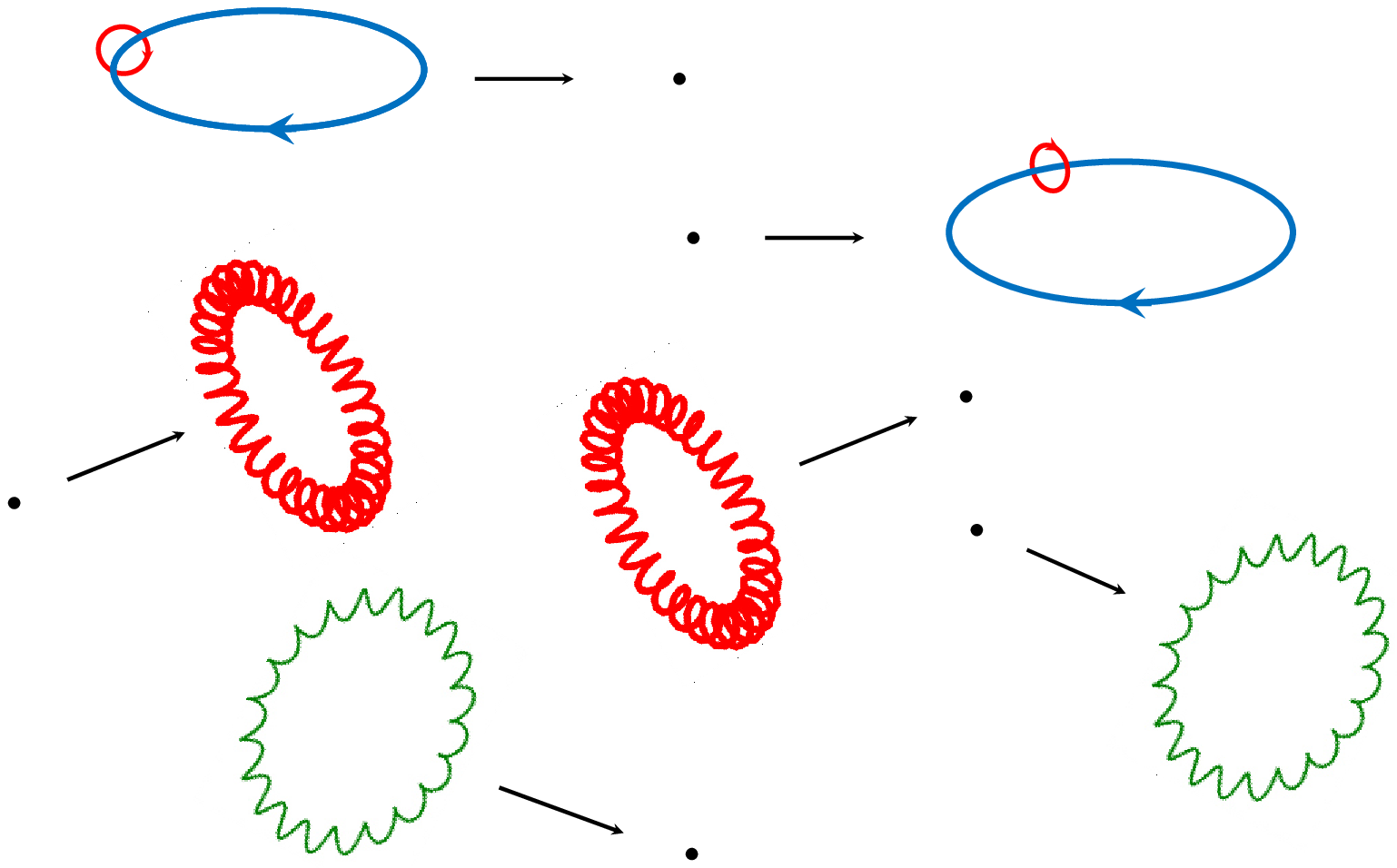}
\vspace{-5mm}
\caption{Processes of the appearance  and disappearance of the vortices.}
\label{Vort_101}
\end{figure}

Thus, the the processes $1 \to 0$ and $0 \to 1$ are perfectly acceptable here (schematically, see Fig.\ref{Vort_101}). The dots in this figure conventionally indicate the disappearance of the vortex rings in a corresponding space point. The different color of the helical rings corresponds to different excited modes ${\mathfrak j}_{-n}$ (see Fig. \ref{Vort_101}).
In a sense, the fluid in which the vortices  in question are formed plays the role of ''aether''.
 { Unlike   Lord Kelvin's  theory, 
   the    ''aether'' in our approach is quite real physical medium  and not hypothetical. 
  Therefore, the processes $1 \to 0$ and $0 \to 1$  do not violate the laws of conservation of energy and momentum: these values can be transmitted both from the vortex to the surrounding vortexless fluid and vice versa. Thus,  energy and momentum will be conserved in the isolated  ''vortices \& fluid'' system. }

According to the author's opinion, it is  such processes that are characteristic of the critical 
mode of a quantum fluid. 
 In classical theory, this mode is characterized by the critical 
Reynolds number ${\mathrm Re_{cr}}$. In a quantum case, the definition of this number is ambiguous 
(see, for example, \cite{HTV,RBAB}).  This problem is outside  of this paper.

Our subsequent purpose is to describe the  quantum states of the system of interacting vortices that allow  processes in question. 
The standard general representation for a density matrix for a quantum statistical system is as follows
\[\hat\rho  ~=~ \sum_k w_k |k\rangle\langle k |\,,\]
where vectors $|k\rangle$ are some pure quantum states. Therefore, we must begin our construction by defining a system of such vectors for the process under consideration. 

To describe the processes $1 \to 0$ and $0 \to 1$ in a system of the evolving vortices, we apply the random Hamiltonians method \cite{Goff}. Let's assume that the operator $\hat{U}$ has the following form for this system in the domain $V$ (see definition (\ref{domain_V})):
\begin{equation}
\label{U_random}
U_{0\leftrightarrow 1} ~=~ i \varepsilon \sum^K_{\langle w\rangle} \bigl(\,{\overline\alpha}_{\langle w\rangle}a^+_{\langle w\rangle} - 
\alpha_{\langle w\rangle}a_{\langle w\rangle}\,\bigr)\,.
\end{equation}
 We have adopted the following notations here:
\begin{enumerate}
\item multi-index $ \langle w\rangle = \{m,\ell, k\,;\, s\,;\, n_1, \dots, n_M \}$, where
the numbers $m,\ell, k$ are the quantum numbers in the space $\boldsymbol{H}_{pq}$, the number $s$ is the quantum number in the space  $\boldsymbol{H}_b$ and the numbers $n_1, \dots, n_M$ are the quantum numbers in the space $\boldsymbol{H}_j$ (see (\ref{space_quant})). 
We assume that the natural number $K$ 
\[ K ~=~  \max\,\{m,\ell, k\,;\, s\,;\, n_1, \dots, n_M \} ~ <~ \infty  \]
exists for the reasons of finiteness of the total energy of our system;
\item creation and  annihilation operators $a^+_{\langle w\rangle}$ and  $a_{\langle w\rangle}$ create and  annihilate the vortex with  set of the quantum numbers $\langle w\rangle$; 
\item the complex numbers $\alpha_{\langle w\rangle}$ are the random numbers such that the condition
$|\alpha_{\langle w\rangle}| = 1$ is fulfilled.
\end{enumerate}

Let's  there are no vortices at the moment $t = 0$, so that $|\Psi(0)\rangle =  |0\rangle$. Therefore,    the evolution of the  quantum   state $|\Psi(t)\rangle$   is defined by the following formula in the interaction representation:  
\begin{equation}
\label{evol_1}
|\Psi(t)\rangle ~=~ e^{\frac{i}{\hbar}  V_{01} t} |0\rangle\,,
\end{equation}
where
\[ V_{01} ~=~ e^{\frac{i}{\hbar}H_0 t}U_{0\leftrightarrow 1}   e^{-\frac{i}{\hbar}H_0 t} \]
Due to the equality  $H_0 |0\rangle = 0$ for the free Hamiltonian $H_0$, the following formula takes place:
\begin{equation}
\label{pure_st}
|\Psi(t)\rangle ~\equiv~ |\Psi(t; \alpha_{\langle w\rangle})\rangle ~=~ 
e^{\frac{i}{\hbar}H_0 t} |t\alpha_{\langle w\rangle}\rangle\,,
\end{equation}
where the states  $|t\alpha_{\langle w\rangle}\rangle$ are the coherent states \cite{Perelomov}
for the finite number of { bosons that are described by }
 the operators  $a^+_{\langle w\rangle}$ and $a_{\langle w\rangle}$.
It is known that the coherent states  can be decomposed through the basis
\[|n_{\langle w\rangle}\rangle ~=~  | m,\ell, k\rangle |s \rangle | n_1, \dots, n_M\rangle\,,
\]
where $ m,\ell, s,\dots, n_M = 0,1,2,\dots\ $ and $k =1,2,\dots$.
According to the considerations above, this decomposition means that the state (\ref{pure_st}) is a superposition of states that correspond to different spatio-temporal scales. 
Therefore, we have  density matrix  discussed in the beginning of this section   in the form:
\[\hat\rho(t)    ~=~ \sum_{\alpha_{\langle w\rangle}} w(\alpha_{\langle w\rangle}) |\Psi(t; \alpha_{\langle w\rangle})\rangle\langle\Psi(t; \alpha_{\langle w\rangle}) |\,,\]
where $w(\alpha_{\langle w\rangle})$ is the uniform distribution of random complex numbers 
$\alpha_{\langle w\rangle}$ on the unit circle.

{ In this section, we have applied  the  developed approach   to describe the interaction of vortices.    The proposed method makes it possible to consider vortex loops as structured bosons quite naturally. One assumption we have made here is that the concentration of the vortices is low.
This assumption allows us to treat vortex creation, annihilation, and interaction as largely independent processes—an acceptable simplification when examining the initial stages of turbulence.    
 A possible method for describing the developed turbulent flow within the framework of the constructed theory was proposed by the author in the work \cite{Tal_24}. 
Probably, the method developed in the paper \cite{Tal_24}  can be considered as some an alternative for  the mean-field   theory (see, for example, \cite{Nayak}). 
 The author aims to explore this topic further  in  subsequent studies.}

\vspace{3mm}

\section{Concluding remarks}

~~~In this study, we have presented the basic statements of the  quantum vortex theory describing the processes of creation and annihilation of isolated single  vortices.
Apparently, the proposed model may be viewed as a  
  simulation of  turbulence emergence in a  quantum  fluid. 
	As for the subsequent applications  to the developed turbulent flows, a modification of the equation (\ref{LIE_pert}) will be necessary \cite{Tal_24}. Moreover,  other terms in the interaction Hamiltonian (\ref{hat_U}), which describe the more complex  processes of decay,  connection and reconnection of vortex rings, must be taken into account.
 We suppose that 
$\varepsilon_{m,n} = \varepsilon_{m,n}(\mathrm Re_q)$. 
 In the proposed approach, an analogue of the Reynolds number $\mathrm Re_q$   for a quantum fluid remains an open question. 

We were also above to  prove that  different quantum states of the model  correspond to the different spatio-temporal scales of the vortices evolution. Such a hierarchy of scales is an additional argument in favor of the possibility of describing quantum turbulent flows within the proposed model.
 The author hopes to return to this
issue in the future.

\end{document}